\documentclass[conference]{IEEEtran}
\onecolumn
\usepackage{setspace}
\usepackage{amsmath,amsfonts}
\usepackage{array}
\usepackage[caption=false,font=normalsize,labelfont=scriptsize,textfont=scriptsize]{subfig} 
\usepackage[colorlinks=true,linkcolor=black,citecolor=black,urlcolor=blue,]{hyperref}
\usepackage{textcomp}
\usepackage{stfloats} 
\usepackage{url}
\usepackage{verbatim}
\usepackage{graphicx}
\usepackage{enumitem}
\usepackage{booktabs} 
\usepackage{bm}
\usepackage{bbding}
\usepackage{mathrsfs} 
\usepackage{amsthm} 
\usepackage{amssymb}
\usepackage{amsmath}
\usepackage{makecell}
\usepackage{cite}

\usepackage{arydshln}
\usepackage{booktabs}
\usepackage{multirow}
\usepackage{tabularx}
\usepackage{tabulary}
\usepackage{indentfirst}
\usepackage{bm}
\usepackage{enumitem}
\setlist[enumerate]{leftmargin=*}
\usepackage[linesnumbered,ruled,vlined]{algorithm2e}
\theoremstyle{remark}
 
\usepackage{colortbl}

\begin{document}
\title{ Stacked Intelligent Metasurface Enabled Near-Field Multiuser Beamfocusing in the Wave Domain}

\author{
 \IEEEauthorblockN{
 Xing Jia\IEEEauthorrefmark{1}\IEEEauthorrefmark{2}, 
 Jiancheng An\IEEEauthorrefmark{3}, 
 Hao Liu\IEEEauthorrefmark{1}, 
 Lu Gan\IEEEauthorrefmark{1}\IEEEauthorrefmark{2},
 Marco Di Renzo\IEEEauthorrefmark{4},
 M\'{e}rouane Debbah\IEEEauthorrefmark{5}, and
 Chau Yuen\IEEEauthorrefmark{3}
 }\\
 \IEEEauthorblockA{
 \IEEEauthorrefmark{1}School of Information and Communication Engineering, University of Electronic Science and Technology of China\\ (UESTC), Chengdu 611731, China\\
 \IEEEauthorrefmark{2} Yibin Institute of UESTC, Yibin 644000, China\\
 \IEEEauthorrefmark{3}School of Electrical and Electronics Engineering, Nanyang Technological University, Singapore 639798, Singapore\\
 \IEEEauthorrefmark{4}Universit\'{e} Paris-Saclay, CNRS, CentraleSup\'{e}lec, Laboratoire des Signaux et Syst\'{e}mes, 91192 Gif-sur-Yvette, France\\
 \IEEEauthorrefmark{5}KU 6G Research Center, Khalifa University of Science and Technology, P O Box 127788, Abu Dhabi, UAE\\
 Email: jiancheng\_an@163.com
 \IEEEauthorrefmark{3} Corresponding author: Jiancheng An
 }
}

\maketitle

\begin{abstract}
Intelligent surfaces represent a breakthrough technology capable of customizing the wireless channel cost-effectively. However, the existing works generally focus on planar wavefront, neglecting near-field spherical wavefront characteristics caused by large array aperture and high operation frequencies in the terahertz (THz). Additionally, the single-layer reconfigurable intelligent surface (RIS) lacks the signal processing ability to mitigate the computational complexity at the base station (BS). To address this issue, we introduce a novel stacked intelligent metasurfaces (SIM) comprised of an array of programmable metasurface layers. The SIM aims to substitute conventional digital baseband architecture to execute computing tasks with ultra-low processing delay, albeit with a reduced number of radio-frequency (RF) chains and low-resolution digital-to-analog converters. In this paper, we present a SIM-aided multiuser multiple-input single-output (MU-MISO) near-field system, where the SIM is integrated into the BS to perform beamfocusing in the wave domain and customize an end-to-end channel with minimized inter-user interference.
Finally, the numerical results demonstrate that near-field communication achieves superior spatial gain over the far-field, and the SIM effectively suppresses inter-user interference as the wireless signals propagate through it.
\end{abstract}


\section{Introduction}\label{Sec_I}
\IEEEPARstart {E}{xtremely} large-scale antenna arrays (ELAAs) and terahertz (THz) communications are promising technologies for sixth-generation (6G) to achieve ultra-high energy and spectrum efficiency \cite{TAP_2024_An_Emerging, ELAAs, VTC_THz, WC_2024_An_Near}, which, however, caused a significant extension of the near-field region, characterized by spherical wavefront \cite{NF_ZengYong, NF_CB_LDMA}. Note that the near-field operation enables higher spatial degrees of freedom (DoF) and new wavefronts over the far-field \cite{NF_Liu_survey}. However, it presents channel modeling, beamfocusing, and hardware deployment challenges. 

Accurate channel modeling is essential for fully harnessing the potential of near-field communications (NFC) \cite{Group_AN_Tutorial_I}. Specifically, the authors of \cite{NF_Liu_survey} utilized uniform plane wave, uniform and non-uniform spherical wave to model the far-field, radiating and reactive near-field channels, respectively. Furthermore, a uniform circular array (UCA) at the base station (BS) was adopted to expand the near-field region in \cite{NF_2}. Moreover, the authors of \cite{NF_ZengYong} proposed a half-space three-dimensional (3D) near-field channel model, which considers phase, amplitude, and projected aperture.

Recently, a promising reconfigurable intelligent surface (RIS) technology has emerged, constructing the virtual line-of-sight (LoS) links to reap spatial gain \cite{TWC_2025_An_Flexible, IoTJ_2023_Xu_OTFS, Group_AN_CB_Mag, TCOM_2024_Yu_Environment, TCT_2024_An_Adjustable}. However, the single-layer RIS is inefficient in signal processing \cite{Group_An_SIM_JSAC}. Thus, \textit{Lin et al.} in \cite{SIM_LingXing} proposed an optical diffractive deep neural network relying on a multilayer structure, which harnesses the wave propagation to realize efficient parallel matrix operations. Furthermore, \textit{Liu et al.} \cite{SIM_Cui_Nature_2022} proposed a programmable stacked intelligent metasurface (SIM) capable of performing multiple computing tasks at optical processing speed.

To mitigate the inter-user interference, the near-field multiuser systems typically require fully digital beamfocusing \cite{TCOM_2022_An_Low, NF_Liu_survey, Group_AN_CB_2}, e.g., zero-forcing (ZF), which incurs significant hardware costs and computational complexity \cite{Group_AN_SIM_Survey, WCL_2024_Niu_Stacked, Group_An_SIM_MU, arXiv_2024_Hao_Multi}. To address this issue, the authors of \cite{, Group_An_SIM_MU_TWC} proposed a novel wave-based signal processing paradigm for suppressing multiuser interference and boosting capacity in SIM-aided wireless systems. Specifically, passive SIM can be integrated into the BS to perform wave-based beamforming to substitute the conventional fully digital beamforming. As a result, only low-resolution digital-to-analog converters (DAC) and fewer radio-frequency (RF) chains are required at the transceiver, leading to substantial cost reductions \cite{WCL_2024_Yao_Channel, WCL_2024_Lin_Stacked}. Additionally, SIM performs forward propagation at the speed of light without incurring extra processing delay \cite{WCL_2024_Huang_Stacked, arXiv_2024_Hao_Multi}. Specifically, \textit{An et al.} \cite{Group_An_SIM_JSAC} optimized SIM-aided holographic point-to-point MIMO system for maximizing the channel capacity. Furthermore, in \cite{Group_An_SIM_MU} and \cite{Group_An_SIM_MU_TWC}, SIM was integrated into the BS to eliminate the inter-user interference to maximize the system sum rate by considering continuous and discrete SIM phase shifts, respectively.

Nevertheless, the SIM in \cite{Group_An_SIM_JSAC, arXiv_2024_Hao_Multi} required an extra power amplifier factor to compensate for the penetration loss. Moreover, the near-field effect due to the large array aperture of SIM was neglected, as well as the practical impedance response of each meta-atom in SIM. In this paper, we study a SIM-aided multiuser multiple-input single-output (MU-MISO) NFC system for joint computing and communication while considering the diffraction behavior of meta-atoms.

\section{ Proposed SIM-Aided Near-Field System Model}\label{Sec_II}
\subsection{The Proposed SIM Model}
\begin{figure}[!t]
\centering
\includegraphics[width=16cm]{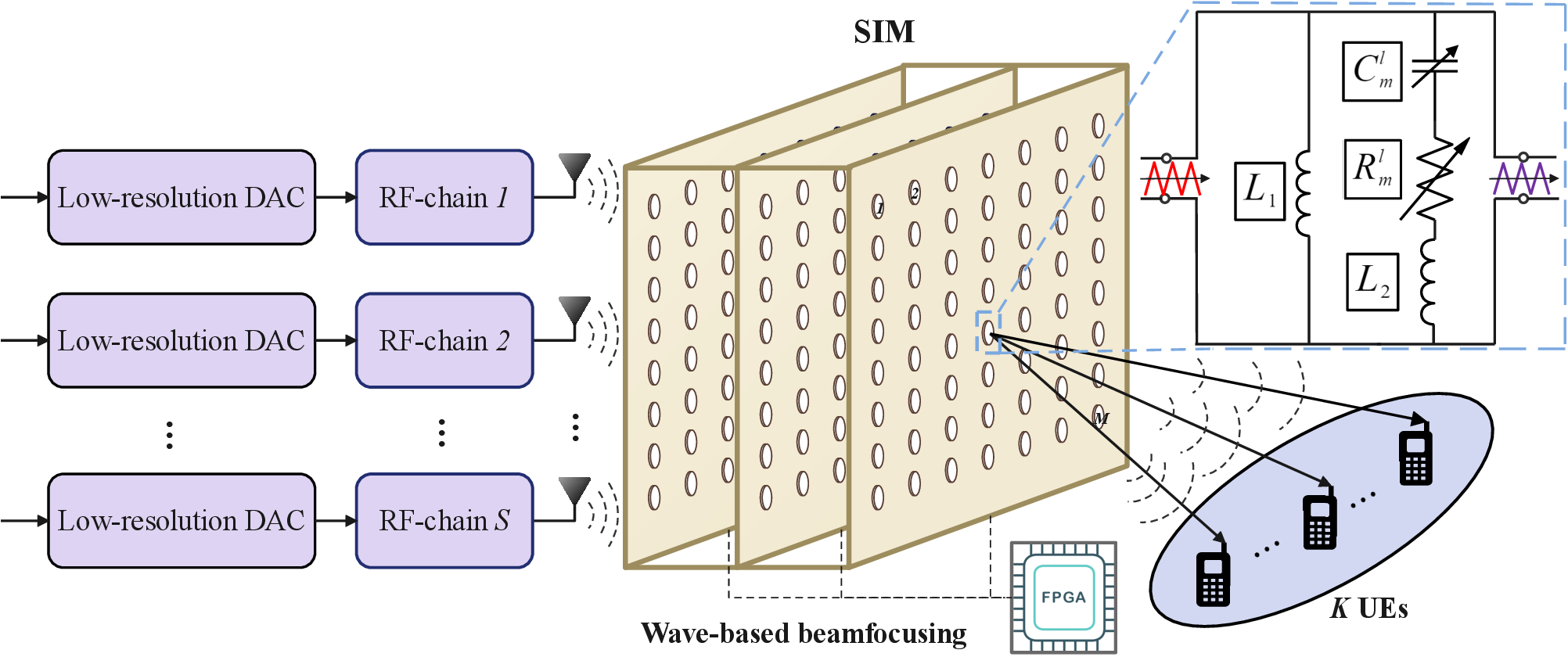}
\caption{A SIM-aided MU-MISO NFC system.}
\label{fig_1}
\end{figure}

As shown in Fig. \ref{fig_1}, we consider a SIM-aided MU-MISO NFC system, where the SIM is integrated into the BS with $S$ antennas to serve $K$ single-antenna user equipments (UEs), satisfying $S=K$\footnote{In contrast to traditional digital baseband architecture, which customizes a beamfocusing vector for each data stream, the innovative SIM-enabled signal processing architecture transmits data streams directly from corresponding transmit antennas. Thus, the BS has to activate a number of antennas that match the number of users for end-to-end multiuser communication.}. The SIM is fabricated by stacking $L$ programmable metasurface layers, and each layer consists of $M$ low-cost passive meta-atoms. Moreover, the SIM is connected to a field-programmable gate array (FPGA) that can independently manipulate the parameters of the SIM meta-atoms. Let $\mathcal{S}= \left\{ 1, 2, \cdots, S\right\}$, $\mathcal{K}= \left\{ 1, 2, \cdots, K\right\}$, $\mathcal{L}= \left\{ 1, 2, \cdots, L\right\}$, and $\mathcal{M}= \left\{ 1, 2, \cdots, M\right\}$ denote the sets of BS antenna, UEs, metasurfaces, and meta-atoms on each metasurface layer, respectively. In addition, the diffraction coefficient of the $m$-th meta-atom on the $l$-th metasurface layer is ${\phi_{m}^{l}}=a_{m}^{l} e^{j{\theta}_{m}^{l}}$, where $a_{m}^{l}$ and ${\theta}_{m}^{l}$ represent the corresponding magnitude and phase responses, respectively, satisfying 
$a_{m}^{l}\in[0,1]$ and ${\theta}_{m}^{l}\in[0, 2\pi)$, $\forall m\in \mathcal{M}, l\in \mathcal{L}$. 
According to \cite{Ref_wu_Z0}, an equivalent model of the impedance of the $m$-th meta-atom on the $l$-th layer can be represented as
\begin{equation}
Z_{m}^{l}\left(C_{m}^{l}, R_{m}^{l}\right)=\frac{\jmath \omega L_{1}\left(\jmath \omega L_{2}+\frac{1}{\jmath \omega C_{m}^{l}}+R_{m}^{l}\right)}{\jmath \omega L_{1}+\left(\jmath \omega L_{2}+\frac{1}{\jmath \omega C_{m}^{l}}+R_{m}^{l}\right)},
\end{equation}
where $C_{m}^{l}$, $R_{m}^{l}$, $L_{1}$, and $L_{2}$ denote the capacitance, resistance, bottom inductance, and top inductance of the $m$-th meta-atom on the $l$-th layer, respectively. 
Moreover, $w$ denotes the angular frequency of the incident electromagnetic (EM) waves. 
Thus, the diffraction coefficient of the $m$-th meta-atom on the $l$-th metasurface layer can be denoted as
\begin{equation}
\Gamma\left(C_{m}^{l}, R_{m}^{l}\right)=\frac{Z_{m}^{l}\left(C_{m}^{l}, R_{m}^{l}\right)-Z_{0}}{Z_{m}^{l}\left(C_{m}^{l}, R_{m}^{l}\right)+Z_{0}},
\end{equation} where $Z_{0}\approx 376.73 \Omega$ is the free space impedance \cite{Ref_wu_Z0}. Thus, we have $a_{m}^{l}=\left | \Gamma\left(C_{m}^{l}, R_{m}^{l}\right) \right | $ and ${\theta}_{m}^{l}=\angle \Gamma\left(C_{m}^{l}, R_{m}^{l}\right)$, $\forall m\in \mathcal{M}, l\in \mathcal{L}$. Furthermore, based on \cite{Ref_wu_Z0}, the coupled behavior between $a_{m}^{l}$ and ${\theta}_{m}^{l}$ can be further simplified by
\begin{equation}\label{ap_model}
a_{m}^{l}\left( {\theta}_{m}^{l} \right)= \left(1-a_\text{min}\right) 
\left( \frac{\sin \left( {\theta}_{m}^{l} -\vartheta \right)+1 }{2} \right)^{\iota }+a_\text{min},
\end{equation} where constant $a_\text{min}$, $\vartheta$, and $\iota$ denote the minimum amplitude, $\left|a_\text{min}-\pi/2\right|$, and curvature of function, respectively.

Ultimately, the diffraction coefficient matrix of the $l$-th metasurface layer can be expressed as $\mathbf{\Phi}^{l}=\text{diag}\left( {\phi_{1}^{l}}, {\phi_{2}^{l}},\cdots,{\phi_{M}^{l}}\right)$, where ${\phi_{m}^{l}} =a_{m}^{l}( {\theta}_{m}^{l})e^{j{\theta}_{m}^{l}} $.
Furthermore, let $\mathbf{W}^{l}\in\mathbb{C}^{M\times M}$, $\forall l \in \mathcal{L} /\{1\}$, and $\mathbf{W}^{1}\in \mathbb{C}^{M\times S}$ denote the propagation coefficient matrix from the $(l-1)$-th layer to the $l$-th metasurface layer and from the BS antenna array to the first metasurface layer, respectively.
Specifically, the $(m,m')$-th item of $\mathbf{W}^{l}$ can be expressed as \cite{Group_An_SIM_JSAC}
\begin{equation}\label{RSDT}
{w_{m, m'}^{l}=\frac{\cos(\psi_{m,m'}^{l}) S_{a}}{r_{m,m'}^{l}}\left(\frac{1}{2 \pi r_{m,m'}^{l}}-j \frac{1}{\lambda}\right) e^{j 2 \pi {r_{m,m'}^{l}} / \lambda} },
\end{equation}
where $\lambda$, $S_{a}$, $\psi_{m, m'}^{l}$, and $r_{m, m'}^{l}$ denote the signal wavelength, the meta-atom area, the angle between the propagation direction and normal direction of the $(l-1)$-th metasurface layer, and the distance between the $m$-th meta-atom on the $l$-th layer and the $m'$-th meta-atom on the $(l-1)$-th layer, respectively.
Similarly, the $(m,s)$-th item of $\mathbf{W}^{1}$ can be derived.

Note that an end-to-end interference-free channel is constructed thanks to the multiuser beamfocusing in the wave domain. Hence, the BS transmits each data stream directly to the corresponding UE, in contrast to the traditional digital baseband system, which transmits a superimposed signal comprising multiple data streams from each antenna.
Thus, the response of the SIM to the propagated EM waves can be expressed as
\begin{equation}\label{SIM_Beamfocusing_matrix}
\mathbf{G}=\boldsymbol{\Phi}^{L} \mathbf{W}^{L} \cdots \boldsymbol{\Phi}^{2} \mathbf{W}^{2} \boldsymbol{\Phi}^{1} \mathbf{W}^{1} \in \mathbb{C}^{M \times K}.
\end{equation}

\subsection{Near-Field Channel Model}\label{II_1}
Next, we adopt a general spherical wavefront to model near-field channel ${h}_{m,k}$ from the $k$-th UE to the $m$-th meta-atom on the $L$-th metasurface layer, which can be denoted by 
\begin{equation}
{h}_{m,k}=e^{-j \frac{2 \pi}{\lambda}{{d}_{m,k}}},
\end{equation}
where ${d}_{m,k}$ denote the distance between the $k$-th UE and the $m$-th meta-atom on the $L$-th layer. 
Furthermore, let $\mathbf{H}=\left[\mathbf{h}_{1}, \cdots,\mathbf{h}_{K} \right] \in \mathbb{C}^{M \times K}$ denote the channel from the $K$ UEs to the $L$-th layer, where $\mathbf{h}_{k}=\left[{h}_{1,k},\cdots,{h}_{M,k} \right]^{T}$.

\subsection{Signal Model for SIM-Aided NFC}
Next, we elaborate on SIM-aided downlink NFC.
Specifically, the end-to-end downlink channel from the BS to $K$ UEs can be represented as
\begin{equation}
\mathbf{Q}=\mathbf{H}^{H}\mathbf{G}\in \mathbb{C}^{K \times K},
\end{equation}
where $\mathbf{Q}=\left[\mathbf{q}_{1},\cdots,\mathbf{q}_{K} \right]$ and $\mathbf{q}_{k}= \mathbf{h}^{H}_{k}\mathbf{G}$ denotes the end-to-end channel from the BS to the $k$-th UE.

Furthermore, let $\mathbf{x}=[{x}_{1},{x}_{2},\cdots,{x}_{K}]^{T}$ denote the transmit symbol vector, where ${x}_{k}$ denotes the data symbol of the $k$-th UE,
and ${x}_{k}$ is an independent random variable with zero mean and unit 
variance, i.e., ${x}_{k}\sim\mathcal{CN}({0},1)$. Thus, the signal received at the $k$-th UE can be denoted as 
\begin{align}\label{down_comm}
{y}_{k} &=\mathbf{h}^{H}_{k}\mathbf{G}\operatorname{diag}\left(\mathbf{p}_{b}\right){\mathbf{x} }+ {n}_{k}, \notag\\
&={\mathbf{h}^{H}_{k}{\mathbf{g}_{k}} \sqrt{{p}_{k}}{{x}_{k} }}+ { {\textstyle \sum_{k' \neq k}^{K}} \sqrt{p_{k'}} \mathbf{h}^{H}_{k} {\mathbf{g}_{k'}} x_{k'}} +{{n}_{k}} ,
\end{align}
where $\mathbf{p}_{b}=[{p}_{1},{p}_{2},\cdots, {p}_{K}]^T$ with $p_{k}$ denoting power allocated to the $k$-th UE, satisfying total power constraint ${P}_{\text{T}}= {\textstyle \sum_{k=1}^{K}}{p}_{k}$.
Moreover, ${\mathbf{g}_{k}}$ denotes the $k$-th column of $\mathbf{G}$ and
${n}_{k}$ denotes the additive white Gaussian noise (AWGN) at the $k$-th UE with the average noise power of ${\sigma^2_{k}}$, satisfying ${{n}_{k}} \sim\mathcal{CN}({0},{\sigma^2_{k}})$. 
Based on (\ref{down_comm}), the signal-to-interference-plus-noise ratio (SINR) of the $k$-th UE can be expressed as 
\begin{equation}\label{SINR}
\gamma_{k}=\frac{{{p}_{k}}\left | \mathbf{h}^{H}_{k}{\mathbf{g}_{k}} \right |^2 }
{\sum_{k'\neq k}^{K} {p_{k'}}\left | \mathbf{h}^{H}_{k} {\mathbf{g}_{k'}}\right |^2 +\sigma^2_{k}}.
\end{equation}
Furthermore, the sum rate of $K$ UEs can be expressed by
\begin{equation}
R\left(\mathbf{p}_{b}, \mathbf{\Phi}^{1},\cdots,\mathbf{\Phi}^{L} \right)= {\textstyle \sum_{k=1}^{K}}\log_{2}\left(1+\gamma_{k} \right).
\end{equation}

\section{Wave-based Beamfocusing Scheme}\label{Sec_III}
\subsection{Problem Formulation}
In this paper, we aim to maximize the system sum rate for the NFC system by optimizing $\mathbf{p}_{b}$ and $\mathbf{\Phi}^{l}, {\forall}{l}\in\mathcal{L}$, which are intricately coupled.
Next, we reformulate the problem to minimize the normalized mean square error (NMSE) between the SIM-based end-to-end channel matrix and the target interference-free one, which facilitates the decoupling of $\mathbf{p}_{b}$. The specific optimization problem is formulated as 
\begin{subequations}\begin{align} 
\mathscr{P}1: \underset{ {\phi_{m}^{l}}} {\text{min}}& \quad {\varpi }= \frac{\left \| {\mathbf{H}^{H}\mathbf{G}} - \mathbf{H}^{H}\mathbf{W}_{\text{ZF}} \right \|_{F}^{2} }{\left \|\mathbf{H}^{H}\mathbf{W}_{\text{ZF}} \right \|_{F}^{2}} \label{min} \\
\text{s.t.} \quad & \mathbf{G}=\boldsymbol{\Phi}^{L} \mathbf{W}^{L} \cdots \boldsymbol{\Phi}^{2} \mathbf{W}^{2} \boldsymbol{\Phi}^{1} \mathbf{W}^{1}, \label{min_b} \\
 & \mathbf{\Phi}^{l}=\text{diag}\left( {\phi_{1}^{l}}, {\phi_{2}^{l}},\cdots,{\phi_{M}^{l}}\right), \forall l\in \mathcal{L}, \label{min_c} \\
 & {\phi_{m}^{l}} =a_{m}^{l}( {\theta}_{m}^{l}) e^{j{\theta}_{m}^{l}}, \ \forall l\in \mathcal{L}, m\in\mathcal{M}, \label{min_d} \\
 & {\theta}_{m}^{l}\in [0,2\pi),\ \forall l\in \mathcal{L},m\in \mathcal{M}, \label{min_e} 
\end{align}\end{subequations}
where $\mathbf{W}_{\text{ZF}} =\mathbf{H} \left({\mathbf{H}^{H}} {\mathbf{H}} \right)^{-1}$ is ZF beamfocusing matrix relying on fully digital baseband for eliminating intra-cell interference among UEs, which incurs the high hardware cost and computational complexity, we utilize the SIM to substitute digital baseband to perform wave-based beamfocusing.
However, due to the non-convex constraint of phase shifts, as well as the fact that the phase and amplitude response of each meta-atom are deeply coupled, solving $\mathscr{P}1$ is non-trivial.

\subsection{Proposed Diffraction Coefficient Optimization Algorithm}
To solve the non-convex problem $\mathscr{P}1$, we propose a gradient descent algorithm to iteratively compute the partial derivatives of ${\varpi}$ with respect to ${\theta}_{m}^{l}$ and regularize the partial derivatives to mitigate issues of gradient explosion and vanishing, as well as compute corresponding ${a}_{m}^{l}(e^{j\theta})$.
The specific optimization algorithm for $\mathscr{P}1$ can be summarised as follows.

After initialising the SIM diffraction coefficient $a_{m}^{l}( {\theta}_{m}^{l}) $ and $ {\theta}_{m}^{l},\forall l\in\mathcal{L},m\in\mathcal{M} $, we calculate the partial derivative of $ {\varpi}$ with respect to ${\theta}_{m}^{l}$ as
\begin{align}\label{gradient_SIM}
&\frac{\partial \varpi}{\partial {\theta}_{m}^{l}} =\frac{2}{\tau} \sum_{k=1}^{K} \sum_{\tilde{k}=1}^{K} \left\{ \Im\left[\left( a_{m}^{l}( {\theta}_{m}^{l}) e^{j{\theta}_{m}^{l}} v_{m, k, \tilde{k}}^{l} \right)^{*}\left( {q_{k, \tilde{k}}}-\Lambda_{k, \tilde{k}}\right)\right] \right. \notag\\
& +\left. \Re \left[\left( \frac{\partial {a_{m}^{l}( {\theta}_{m}^{l})}}{\partial {\theta}_{m}^{l}} e^{j{\theta}_{m}^{l}} v_{m, k, \tilde{k}}^{l} \right)^{*}\left( {q_{k, \tilde{k}}} -\Lambda_{k, \tilde{k}}\right)\right] \right\},
\end{align}
where $\mathbf{\Lambda}=\mathbf{H}^{H}\mathbf{W}_{\text{ZF}}$ and $\tau={\left \| \mathbf{\Lambda} \right \|_{F}^{2}}$,
$q_{k, \tilde{k}}$ and $\Lambda_{k, \tilde{k}}$ denote the $(k, \tilde{k})$-th entry of $\mathbf{Q}$ and $\mathbf{\Lambda}$. 
Moreover, we have
\begin{align}
\frac{\partial {\left( {a_{m}^{l}( {\theta}_{m}^{l})} \right)}^{*} } { \partial { \theta}_{m}^{l}} &=
 \frac {{\iota} \left(1-a_\text{min}\right) }{2} {\cos \left( {\theta}_{m}^{l} -\vartheta \right)} \times \notag\\ 
&\left( {\sin \left( {\theta}_{m}^{l} -\vartheta \right)+1 } \right)^{\iota-1 } \left({e^{j{\theta}_{m}^{l}} 
 {v}_{m, k, \tilde{k}}^{l} } \right)^*,
\end{align}
and $v_{m, k, \tilde{k}}^{l}$ denotes the cascaded channel spanning from the $\tilde{k}$-th BS antenna to the $k$-th UE, which is defined by
\begin{equation}\label{equ_vmkk}
{v}_{m, k, \tilde{k}}^{l}=\mathbf{h}^{H}_{k} \boldsymbol{\Phi}^{L} \mathbf{W}^{L} \cdots \mathbf{W}_{:, m}^{l+1} \mathbf{W}_{m,:}^{l} \cdots \boldsymbol{\Phi}^{1} \mathbf{W}^{1}_{{:, \tilde{k}}},
\end{equation} 
where $\mathbf{W}_{:, m}^{l+1}$, $\mathbf{W}_{m,:}^{l}$, and $\mathbf{W}^{1}_{{:, \tilde{k}}}$ denote the $m$-th column of 
$\mathbf{W}^{l+1}$, the $m$-th row of $\mathbf{W}^{l}$, and the $\tilde{k}$-th column of $\mathbf{W}^{1}$, respectively.
Subsequently, we normalize ${\partial \varpi}/{\partial {\theta}_{m}^{l} }$ as
\begin{equation}\label{gradient_clc}
\frac{\partial \varpi}{\partial {\theta}_{m}^{l} } \leftarrow \frac{\pi}{ \mu_{l}} \cdot \frac{\partial \varpi}{\partial {\theta}_{m}^{l} }, \forall m \in \mathcal{M}, l \in \mathcal{L},
\end{equation}
where $\mu_{l}$ denotes the maximum sensitivity of $\frac{\partial \varpi}{\partial \theta_{m}^{l}}$, $\forall m \in \mathcal{M}$, which can be expressed by 
\begin{equation}\label{Normalization}
\mu_{l}=\max _{m \in \mathcal{M}}\left(\frac{\partial \varpi}{\partial \theta_{m}^{l}}\right), \forall l \in \mathcal{L}.
\end{equation}
Furthermore, we employ a smoothing update strategy as
\begin{equation}\label{Configuring}
{\theta}_{m}^{l} \leftarrow {\theta}_{m}^{l}-\eta \frac{\partial \varpi}{\partial {\theta}_{m}^{l} }, \forall m \in \mathcal{M}, l \in \mathcal{L},
\end{equation}
where $\eta\in(0,1)$ denotes the learning rate. Meanwhile, we update the amplitude response $a_{m}^{l}( {\theta}_{m}^{l})$, $\forall m\in\mathcal{M},l\in\mathcal{L}$ based on (\ref{ap_model}).
Additionally, the learning rate in each iteration is reduced to
\begin{equation}\label{Update}
\eta \leftarrow \rho \eta ,
\end{equation}
where $\rho\in(0,1)$ denotes the decay rate of $\eta$, facilitating fine-grained tuning. Specifically, 
this approach reduces the tuning space as the model nears the optimal solution, preventing oscillation of optimization variables.

Repeatedly executing the above steps, $\varpi$ gradually decreases and converges to its minimum within a moderate number of iterations. Moreover, we use the codebook-based solution \cite{Group_AN_CB_Mag} as the initial solution to reduce the risk of the gradient descent algorithm falling into the locally optimal solutions.

Given the $\mathbf{\Phi}^{l}, \forall l\in \mathcal{L}$, we optimize the BS power allocation vector based on the iterative water-filling (WF) algorithm \cite{Group_An_SIM_JSAC} and the power allocated to the $k$-th UE can be expressed as
\begin{equation}
 {p_{k}}= \left(\kappa- \frac {\textstyle \sum_{k'\neq k}^{K} {p_{k'}}\left | \mathbf{h}^{H}_{k} {\mathbf{g}_{k'}}\right |^2 +\sigma^2_{k}} {{{p}_{k}}\left | \mathbf{h}^{H}_{k}{\mathbf{g}_{k}} \right |^2 }
 \right)^{+}, \forall k \in \mathcal{K}, 
\end{equation}
where $\kappa$ is the water-filling level, determined by binary search, while ${p_{k}}$ satisfies the constraints ${\textstyle \sum_{k=1}^{K}} {p_{k}}=P_\text{T}$ and ${p_{k}} \ge 0, \ {\forall}{k}\in\mathcal{K}$.

\subsection{The Algorithm Complexity Analysis}
Next, we evaluate the computational complexity of the proposed algorithm. Specifically,
the complexity of SIM diffraction coefficient optimization based on the gradient descent algorithm is $\mathcal{O}_\text{GDA}=4I_\text{GDA}\left[ (2L-2)M^3 + MLK^2\right]$, where $I_\text{GDA}$ denotes the number of iterations. Also $4\left[ (2L-2)M^3+2K^2(M+1)+M^2S\right]$, $\mathcal{O}\left( 4MLK^2\right)$, $\mathcal{O}\left( 4ML\right)$, $\mathcal{O}\left( ML\right)$, $\mathcal{O}\left( 1\right)$ denote computational complexity of the forward propagation, gradient calculation, regularization, SIM diffraction coefficient update, and learning rate update, respectively.

\section{Simulation Results}\label{Sec_IV}
This section presents the numerical results to verify the performance of the SIM-aided MU-MISO NFC system.
Specifically, we consider a SIM-aided MU-MISO downlink NFC system with operating frequency $f=10$ GHz, corresponding to the wavelength $\lambda=3$ mm, and the BS is a $2\times 2$ UPA on the $x-z$ plane with antenna spacing $d_{S}=\lambda$. 
The SIM, also on the $x-z$ plane, consisting of $L$ isomorphic metasurface layers, each layer is a $15\times15$ UPA. The layer spacing, meta-atom spacing, and the area of each meta-atom are $d_{L}=1.2\lambda$, $d_{M}=\lambda$, and $S_{a}=d_{M}^2$, respectively.
Unless other specified, we set $S=K=4$, $M=225$, and $L=12$, respectively, 
resulting in the Rayleigh distance of $R_{Ray}=23.52$ m.
Moreover, we set $\rho=0.9$, $\eta=0.99$, and codebook size $T=200$ for proposed and benchmark schemes.
In addition, we set the parameters of the equivalent coupled phase/amplitude model for each meta-atom are $a_\text{min}=0.2$, $\vartheta=0.43 \pi$, and $\iota=1.6$, respectively. The corresponding practical circuit configuration are $L_1=2.5$ nH, $L_2=0.7$ nH, $R_{m}^{l}=2.5 \Omega$, and $0.47 \le C_{m}^{l} \le 2.35$ pF, $\forall m\in\mathcal{M}, l\in\mathcal{L}$, respectively \cite{Ref_wu_Z0}.
Moreover, the BS and SIM are deployed at the height of $H_\text{BS}=H_{S}=3$ m. Moreover, $K$ UEs are randomly distributed within a $r_{u}=3$ m radius near-field on the $x-y$ plane, centered at $(x_{0},y_{0},0)=(3,3,0)$.
Furthermore, path loss is modeled as $\beta = (\lambda/{4\pi})^2 d^{-\alpha} $, where $d$ and $\alpha=2.8$ denote the distance of the link and path loss exponent, respectively. 
Moreover, the transmit power at the BS is ${P_\text{T}} = 5$ dBm and the average noise power at the UE is ${\sigma^2_k} = - 120$ dBm, $\forall k \in \mathcal{K}$. Finally, all simulation results are averaged over $100$ independent experiments.

\begin{figure}[!t]
\centering
\includegraphics[width=8cm]{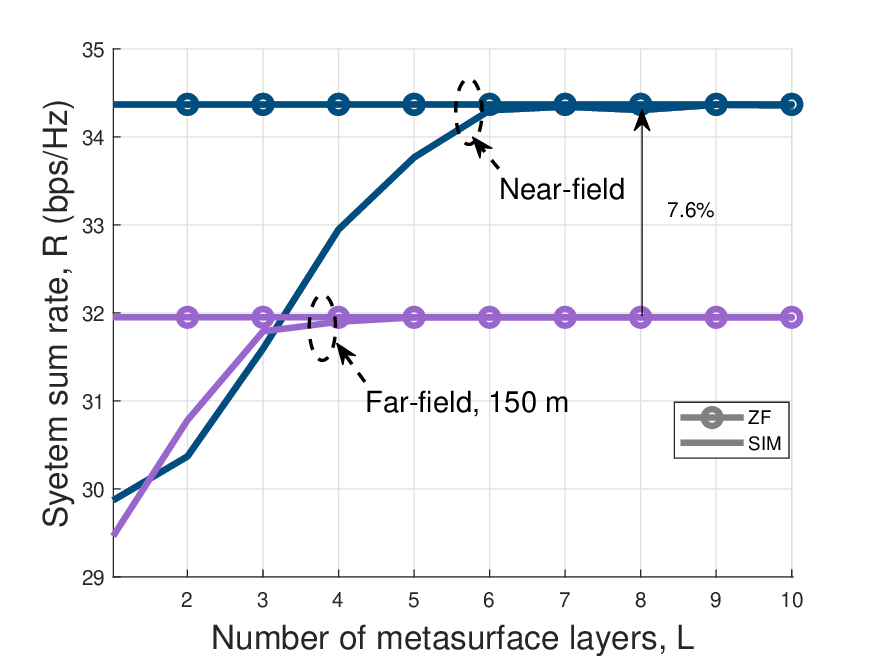}
\caption{The system sum rate versus $L$ under near-field and far-field channel. ($S=K=4$ and $P_\text{T}= -40$ dBm).}
\label{fig_3}
\end{figure}
As shown in Fig. \ref{fig_3}, we evaluate the impact of the far-field \cite{Group_Jia} and near-field channel modeling on the SIM-aided MU-MISO system. 
In order to evaluate the sum rate improvement caused by NFC, we consider the normalized channel gain.
Specifically, Fig. \ref{fig_3} demonstrates that the system sum rate of the near-field channel model is $7.6$\% higher than that of the far-field with $150$ m.
 This is due to the fact that the near-field takes into account both the distance and the angle for the spherical wavefront channel modeling.
As a result, the channel matrix having a higher rank results in a higher sum rate.
Moreover, under all setups, the SIM relying on computation in the wave domain approaches the ZF precoding performance.

\begin{figure}[!t]
\centering
\subfloat[]{\includegraphics[width=8cm]{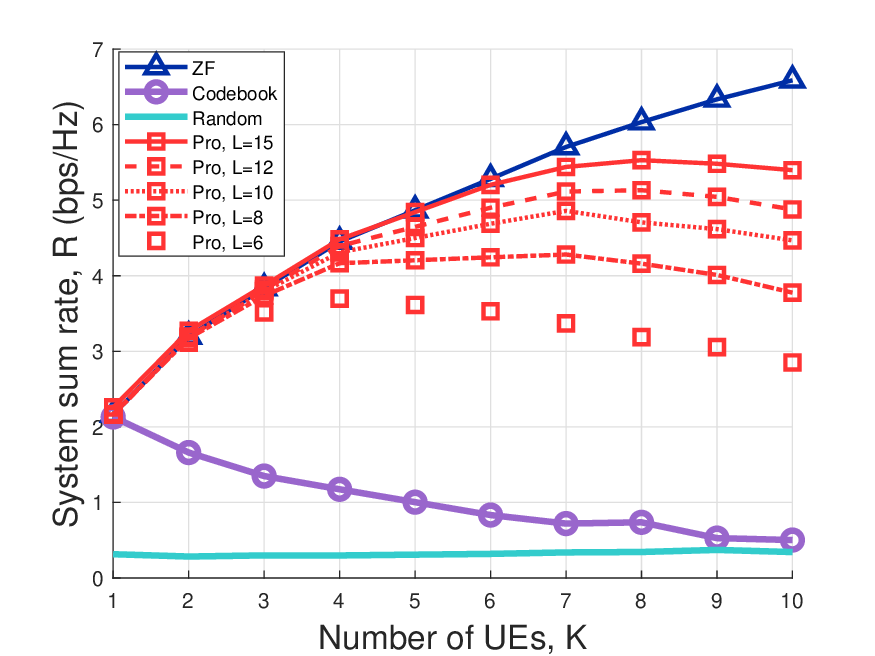}}\\
\subfloat[]{\includegraphics[width=8cm]{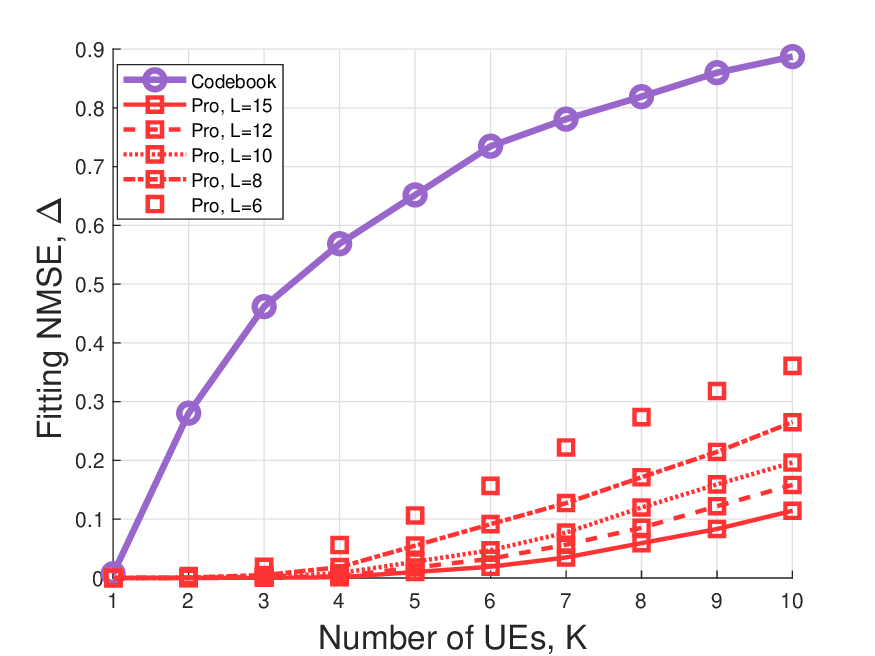}}
\caption{ (a) The system sum rate versus the number of UEs, $K$. (b) The fitting NMSE ${\varpi }$ versus $K$. ($P_\text{T}= 5$ dBm, $M=225$ and $S=K$).}
\label{fig_7}
\end{figure}
Fig. \ref{fig_7}(a) evaluates the system sum rate versus the number of UEs $K$, where we consider codebook and random configuration benchmark schemes \cite{Group_An_SIM_MU_TWC}. 
Note that the performance of the codebook scheme gradually decreases, due to the fact that the codebook-based solution results in the non-ignorable inter-user interference, which becomes severe as $K$ increases.
By contrast, the proposed algorithm effectively utilizes the multiuser multiplexing gain for a small number of $K$, and it may result in a diminished capacity return as $K$ increases beyond a certain value. However, the performance of the proposed scheme can be improved by gradually increasing $L$. 
Fig. \ref{fig_7}(b) portrays the fitting NMSE ${\varpi }$ versus $K$, where ${\varpi }$ of all schemes decrease as $K$ increases. This is because the increasing dimension of $\mathbf{Q}$ makes it more challenging to mitigate inter-user interference using a SIM with the fixed size, thus leading to performance degradation. Note that a higher ${\varpi}$ is obtained as $M$ or $L$ increases, which reveals a fundamental trade-off between the hardware complexity and fitting NMSE.

\begin{figure}[!t]
\centering
\subfloat[]{\includegraphics[width=8cm]{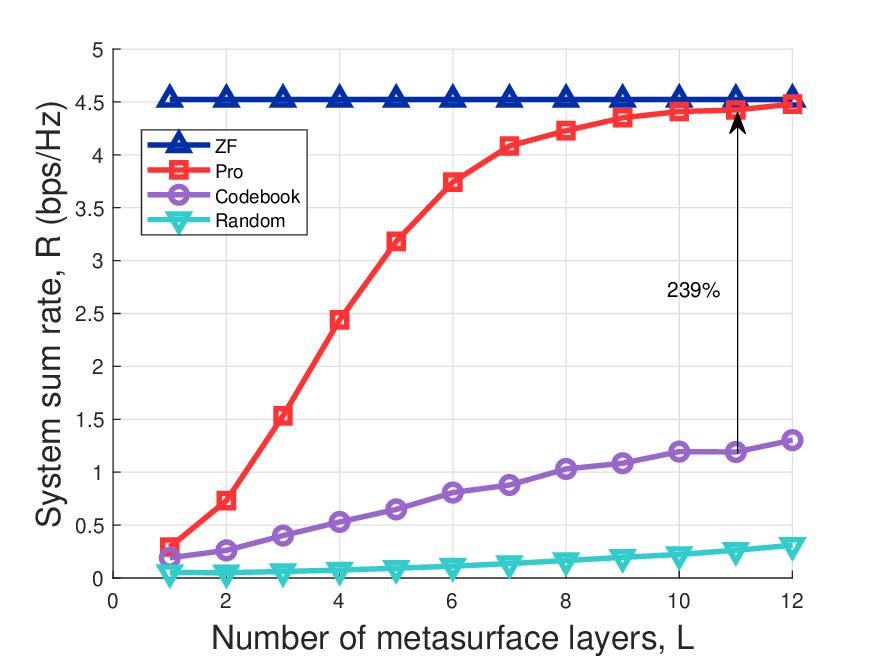}}\\
\subfloat[]{\includegraphics[width=8cm]{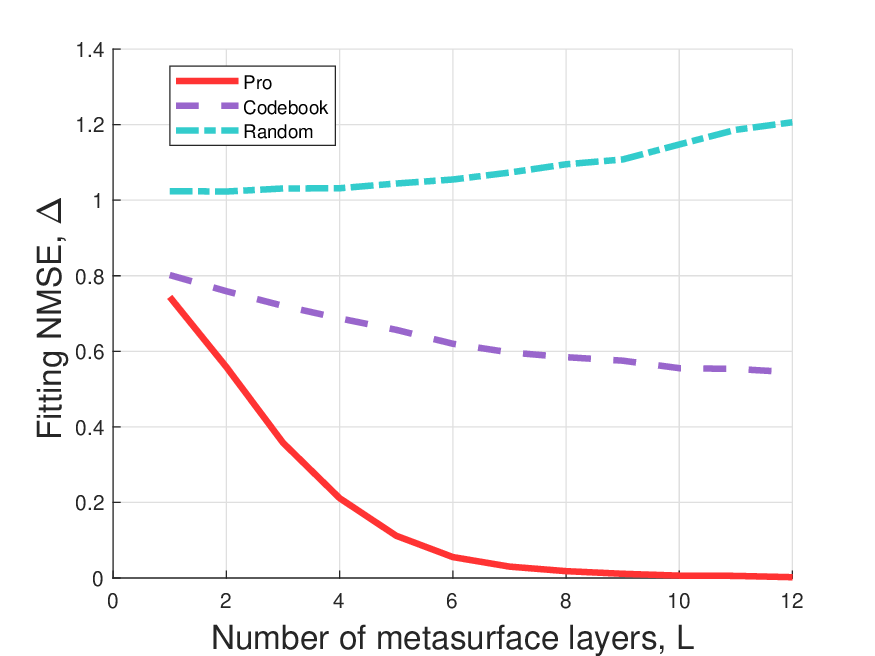}}
\caption{ (a) The system sum rate versus the number of metasurface layers, $L$. (b) The fitting NMSE ${\varpi }$ versus $L$. ($P_\text{T}= 5$ dBm and $S=K=4$).}
\label{fig_8}
\end{figure}
Next, we evaluate the system sum rate versus the number of metasurface layers $L$. 
As shown in Fig. \ref{fig_8}(a), the proposed algorithm achieves $239\%$ higher sum rate than the codebook scheme at $L=11$.
Moreover, the sum rate of SIM-aided NFC system improves rapidly as $L$ increases, thanks to the enhanced signal processing capability brought by multilayer SIM to eliminate inter-user interference.
Fig. \ref{fig_8}(b) portrays the fitting NMSE ${\varpi }$ versus $L$. It is observed that the codebook scheme results in a higher value of $\varpi $ due to the limited codebook size. By contrast, the fitting NMSE of the proposed gradient descent algorithm decreases as $L$ increases. Specifically, ${\varpi}$ decreases from $0.74$ to $0.002$ as $L$ increases from $1$ to $12$, which implies that multilayer SIM has the ability to effectively tune incident EM waves and suppresses the inter-user interference in the wave domain. 

\begin{figure}[!t]
\centering
\subfloat[$L=1$]{\includegraphics[width=8cm]{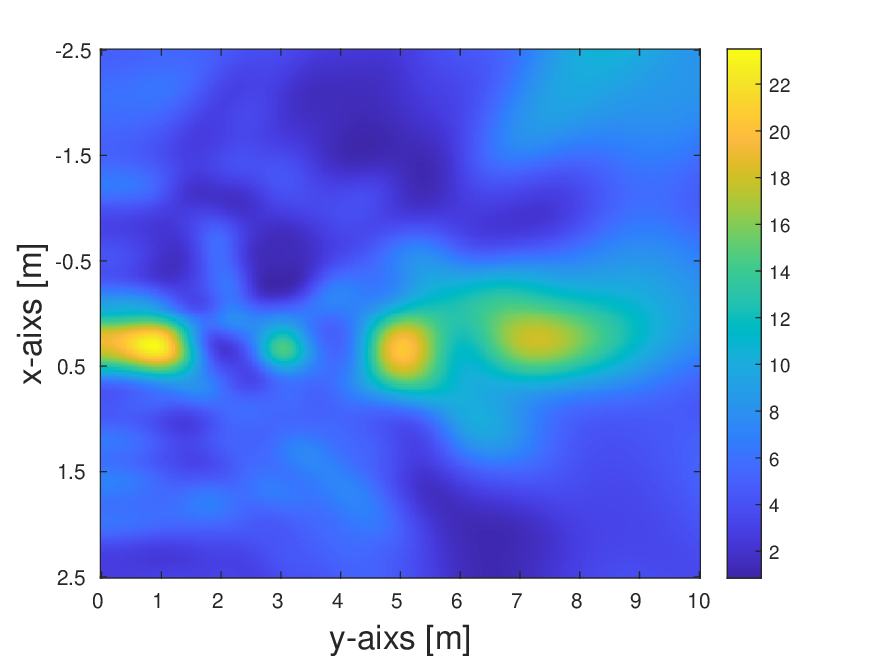}} 
\subfloat[$L=2$]{\includegraphics[width=8cm]{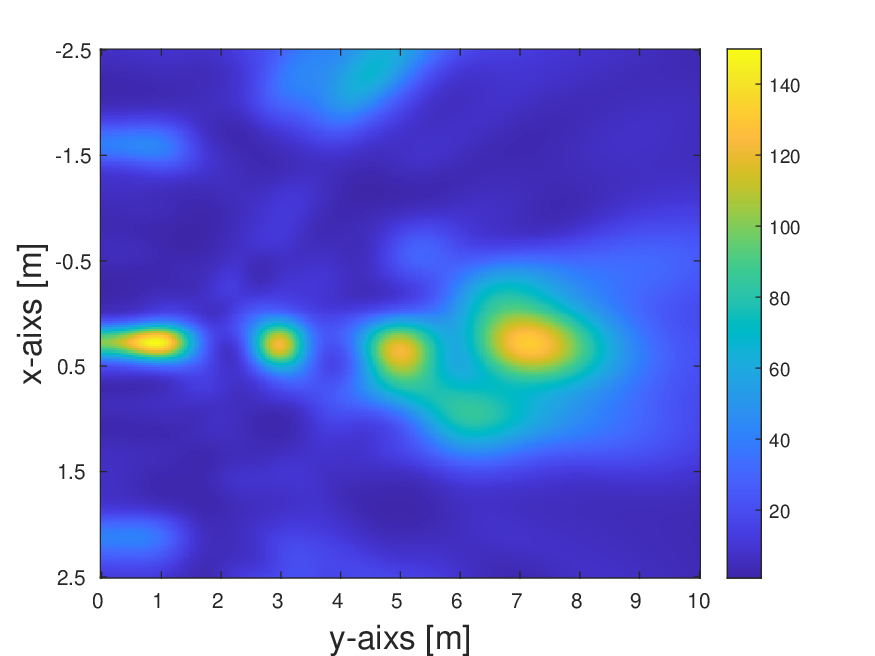}}\\
\subfloat[$L=4$]{\includegraphics[width=8cm]{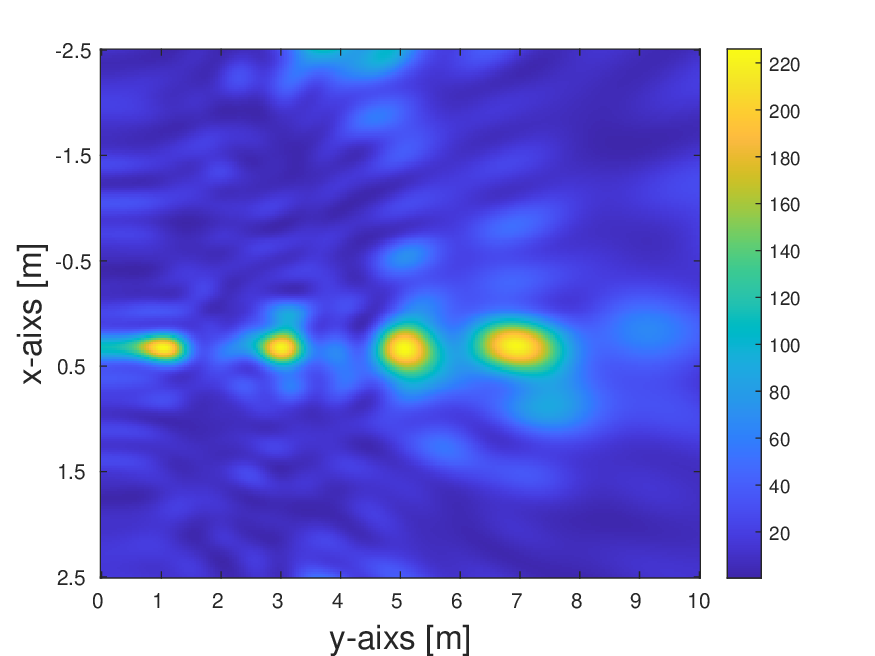}} 
\subfloat[ZF]{\includegraphics[width=8cm]{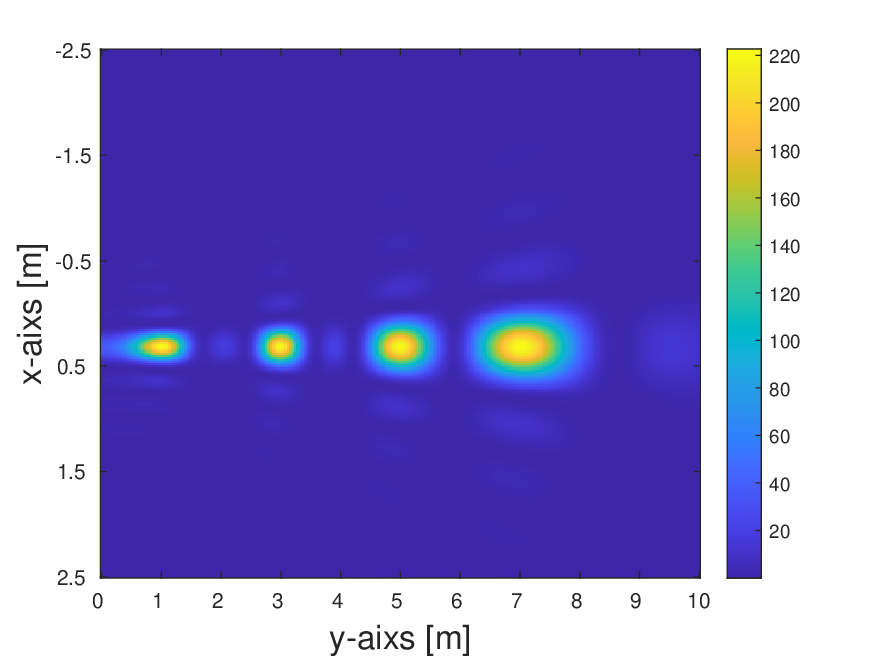}}
\caption {The visualization of the near-field beamfocusing for UEs in the same direction. ($S=K=4$).}
\label{fig_10}
\end{figure}
As shown in Fig. \ref{fig_10}, to verify the ability of SIM beamfocusing to suppress inter-user interference, we visualize the wave-based beamfocusing for the SIM-aided MU-MISO NFC system.
Specifically, as shown in Fig. \ref{fig_10}, we deploy $K$ UEs on the $y$-axis at intervals of $1.5$ m in the same direction, perpendicular to the $x-z$ plane of the SIM.
Then, for a $(5\times10)$ m$^2$ area on the $x-y$ plane, the received energy of the sample point $(x,y,0)$ is calculated based on $ {\textstyle \sum_{k=1}^{K}} \mathbf{h}_{x,y}^{H} \mathbf{g}_{k}$, where $\mathbf{h}_{x,y}^{H}$ denotes the downlink channel from SIM to sample point $(x,y,0)$.
Moreover, Fig. \ref{fig_10}(a) demonstrates that the single-layer metasurface can only focus energy on two UEs, due to its limited tuning capability for the incident EM waves. In contrast, Figs. \ref{fig_10}(b) and (c) demonstrate that further increasing the number of layers to $L=4$ achieves comparable focusing performance with ZF, benefiting from interference cancellation capability brought by SIM.
Moreover, note that the proposed algorithm for wave-based beamfocusing is a suboptimal solution, leading to minor energy leakage.

\section{Conclusion}\label{Sec_V}
In this paper, we developed a SIM-enabled joint computing and communication framework for the MU-MISO NFC system, where the SIM is integrated into the BS to substitute digital beamfocusing architecture with advanced wave-based computation. 
This new computing paradigm relying on wave propagation substantially improves processing speed while simplifying the hardware architecture and computational complexity. 
Specifically, we adopted a general spherical wavefront to characterize the propagation of EM waves for exploiting the high DoF of the near-field channel.
Moreover, we formulated an optimization problem that minimizes the inter-user interference and then proposed an iterative gradient descent algorithm to optimize the SIM diffraction coefficients.
Furthermore, simulation results demonstrated that the SIM in the near-field yields a superior sum rate compared to its far-field counterpart. Finally, it is verified that SIM has the ability to reconstruct an almost perfect end-to-end channel between BS and UEs, thereby effectively eliminating inter-user interference in the wave domain. 
In a nutshell, wave-based beamfocusing provides a novel signal processing paradigm for NFC systems.

\bibliographystyle{IEEEtran}
\bibliography{IEEEabrv,Ref.bib}
\end{document}